\documentclass{ws-procs975x65}
\def\beq{\begin{equation}}
\def\eeq{\end{equation}}
\def\figsubcap#1{\par\noindent\centering\footnotesize(#1)}

\begin{document}

\title{What's the Matter in Cosmology?}
\author{Timothy Clifton}
\address{School of Physics \& Astronomy, Queen Mary University of London,\\
Mile End Road, London E1 4NS, UK\\
E-mail: t.clifton@qmul.ac.uk}

\begin{abstract}
Almost all models of the universe start by assuming that matter fields can be modelled as dust. In the real universe, however, matter is clumped into dense objects that are separated by regions of space that are almost empty. If we are to treat such a distribution of matter as being modelled as a fluid, in some average or coarse-grained sense, then there a number of questions that must be answered. One of the most fundamental of these is whether or not the interaction energy between masses should gravitate. If it does, then a dust-like description may not be sufficient. We would then need to ask how interaction energies should be calculated in cosmology, and how they should appear in the Friedmann-like equations that govern the large-scale behaviour of the universe. I will discuss some recent results that may shed light on these questions. 
\end{abstract}


\bodymatter

\section{Introduction}

The real Universe contains matter fields that are organised into a hierarchy of non-linear structures, ranging from stars and planets, to galaxies and clusters of galaxies. The standard model of cosmology, on the other hand, treats the matter content of the large-scale Universe as being well modelled by a continuous, and (almost) homogeneous fluid of pressureless dust. Non-linear structures are only included as perturbations to a background that has already been determined in this way, and even then only in a Newtonian fashion. This may well be sufficient to model the Universe we see around us, but there are a number of questions that one might like to answer in order to have confidence in this conclusion. Some of these questions involve the consequences of the interaction energies between massive bodies.
\begin{figure}[b!]
\begin{center}
\includegraphics[width=3.1in]{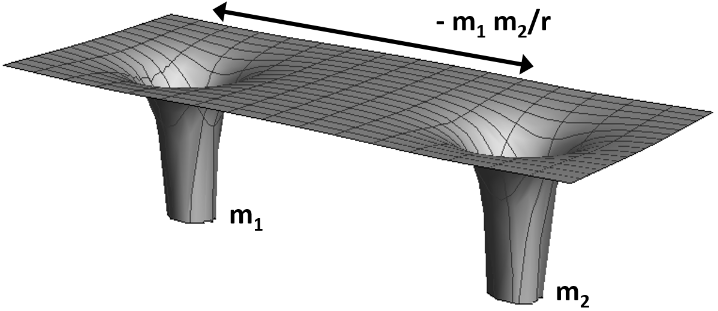}
\caption{Two black holes of mass $m_1$ and $m_2$, separated by distance $r$.}
\label{fig1}
\end{center}
\end{figure}

Let us start by recalling how interaction energies work in relativistic gravity. It is well known that gravity gravitates in Einstein's theory, and one aspect of this statement is the fact that the gravitational potential energies between point-like masses in an asymptotically flat space are themselves a source of gravity\cite{bl}. This situation is illustrated in Fig. \ref{fig1}, where the total mass inferred from observing the gravitational field of this system at infinity is given by $M \simeq m_1+m_2-m_1 m_2/r$. That is, the masses of the two black holes do not add linearly to give the mass of the system as a whole. This curious phenomenon raises the question of whether the interaction energies between masses can affect the large-scale expansion of the Universe, in the same way that mass does. It also raises the question of how one should go about trying to calculate the magnitude of interaction energies in cosmology.

Neither of these questions have straightforward answers. One of the reasons for this is the fact that cosmological models do not usually have asymptotically flat regions, making it difficult to define mass in the first place. A further difficulty arises from the fact that Newtonian-like gravitational potentials may not be well defined over cosmological distances, and could be problematic to sum up if there are infinitely many masses, or if space is topologically non-trivial. What is more, the lack of cosmological models with realistic, gravitationally bound objects means that there are practical difficulties in trying to explore these questions, without already assuming that the large-scale behaviour is given by dust. Nevertheless, we should expect interactions between masses to be present in cosmology. The question is: How do these interactions affect the large-scale behaviour of space?

\section{Cosmological Models with Point-Like Masses}

\begin{figure}[b!]
\begin{center}
  \parbox{2.1in}{\includegraphics[width=2in]{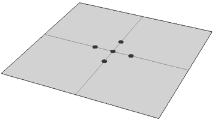}\figsubcap{a}}
  \hspace*{4pt}
  \parbox{1.4in}{\includegraphics[width=1.3in]{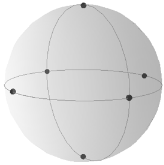}\figsubcap{b}}
  \caption{Two examples of how to position masses in (a) a flat space, and  (b) a spherical space.}
  \label{fig2}
\end{center}
\end{figure}

One can make a start at constructing cosmological models that contain point-like masses, rather than continuous fluids, by considering cosmology as an initial value problem. The solutions to the constraint equations on the initial hyper-surface are then greatly simplified if we take this surface to be instantaneously static. If we position masses in a flat space of this kind, as illustrated in Fig. \ref{fig2}(a), then it can be seen that the constraints are solved by\cite{misner}
\beq
dl^2 = \psi^4 (dx^2+dy^2+dz^2) \hspace{20pt} {\rm where} \hspace{20pt} \psi = 1 + \sum_{i} \frac{\tilde{m}_i}{r} \, ,
\eeq
where $\tilde{m}_i$ are a set of constant parameters. They contain information about both the proper mass of each source, as well as the interaction energies between sources. If, on the other hand, we choose to position our masses on a spherical space, as illustrated in Fig. \ref{fig2}(b), then we find that a solution is given by\cite{paper1}
\beq
dl^2 = \psi^4 (d \chi^2 + \sin^2 \chi d \Omega^2) \hspace{20pt} {\rm where} \hspace{20pt} \psi = \sum_{i} \frac{\sqrt{\tilde{m}_i}}{2 \sin (\chi_i/2)} \, ,
\eeq
where the $\tilde{m}_i$ parameters are again a set of constants. This latter situation looks a bit more like a cosmological model. It also allows a great deal of freedom, as the masses can be positioned arbitrarily, and the $\tilde{m}_i$ parameters can be chosen at will.

\section{Interactions Between Clustered Masses}

\begin{figure}[h!]
\begin{center}
\includegraphics[width=1.5in]{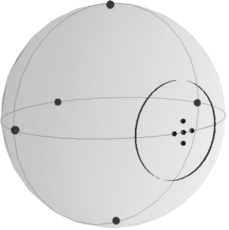}
\caption{A cluster of masses, which are the images of the other (regularly arranged) masses.}
\label{fig3}
\end{center}
\end{figure}

The first, and easiest way that one might try and calculate interaction energies in cosmology is to arrange for there to be clusters of point-like masses. This situation is illustrated in Fig. \ref{fig3}, where a cluster is created by taking the images of what would otherwise be a set of regularly arranged masses\cite{paper2}. In this case there is a relatively natural way to calculate the interactions between the clustered masses, and one finds that the total energy (mass plus interactions) is given by\cite{paper2} $\sum_i \sqrt{\tilde{m}_i}$. This follows from treating the clustered masses as if they exist in a background created by all of the other masses in the universe.
\vspace{-5pt}
\vspace{-10pt}

With a definition in hand, one can now try and determine whether or not the interaction energies between the clustered masses has any effect on the scale of the cosmology. In turns out that they do, and that, in the models constructed in Ref. \refcite{paper2}, their effect on the cosmology is almost exactly the same as the effect of mass (to an accuracy of around $1$ part in $10^8$). This demonstration that interaction energies can effect cosmology is interesting, but it is only a part of the whole problem. To go further we need to consider the effect of the interactions between the un-clustered masses. The remainder of this article will be aimed at doing just this.

\section{Total Interaction Energy}

Let us begin by considering the total interaction energy within the two contructions illustrated in Fig. \ref{fig2}.

\begin{figure}[h!]
\begin{center}
\includegraphics[width=2.5in]{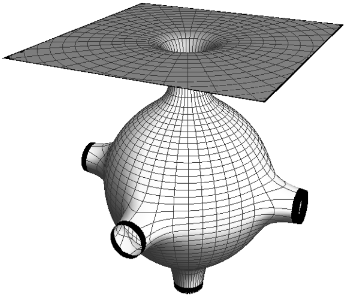}
\caption{An example of the type of cosmological space that can result from strategically positioning masses in the conformal spaces shown in Fig. \ref{fig2}.}
\label{fig4}
\end{center}
\end{figure}

{\bf Interaction Energies in $R^3$:} A cosmological model can be created\cite{lw} by strategically placing masses in the conformal Euclidean space displayed in Fig. \ref{fig2}(a). An example to illustrate the type of space that can result is shown in Fig. \ref{fig4}. The flat region at the top of the image is isometric to the original conformal space, in the asymptotically distant region. Each of the masses we position in the conformal Euclidean space creates an apparent horizon in the bulbous cosmological region below. These are illustrated by thick bands. On the other side of each apparent horizon, and supressed from the image, are a set of extra asymptotically flat regions (one for each mass). The remaining horizon, between the upper sheet and the lower cosmological region, can be though of as the joint horizon of all of the points combined.
\vspace{-5pt}
\vspace{-10pt}

Interaction energies in the Euclidean space are straightforward to calculate\cite{bl}. The total energy in this case (mass plus interactions) is given simply by $\sum_i \tilde{m}_i$, where the sum ranges over the points positioned in the conformal Euclidean space (one fewer than the number of horizons that exist in the cosmological region). If we choose the $\tilde{m}_i$ parameters such that all of the black holes in the cosmological region are identical, then the proper mass of any one of these will be given by the solution of $m_i \simeq (n-1) m_i + E$, where $n$ is the number of black holes and $E$ is the total interaction energy between the points we positioned in the conformal Euclidean space. The ratio of total proper mass to interaction energy can then be seen to be given by $2/n-1$.

\begin{figure}[h!]
\begin{center}
\includegraphics[width=2.6in]{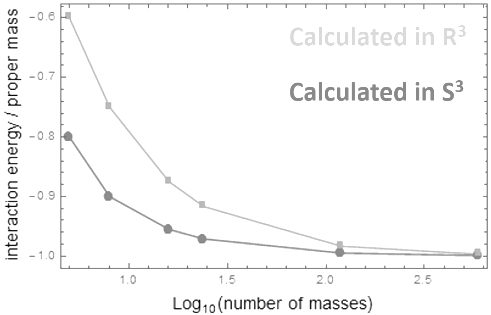}
\caption{Interaction energy as a fraction of total proper mass, for cosmological models with regularly arranged identical black holes.}
\label{fig5}
\end{center}
\end{figure}

{\bf Interaction Energies in $S^3$:} A similar cosmological model can be generated by placing masses regularly in the spherical conformal space illustrated in Fig. \ref{fig2}(b). If the $\tilde{m}_i$ are chosen correctly, so that they are all equal, the cosmological space that results will be the same as that illustrated in Fig. \ref{fig4}. In this case it is more difficult to define exactly what the interaction energy between any two points should be, as there is no asyptotic region from which to observe the gravitational field. The most natural definition would appear to be given by proceeding in formal analogy with the asymptotically flat case, and taking the total energy (mass plus interactions) to be given by $\sum_i \tilde{m}_i$. The sum of the proper masses of all of the black holes can then be subtracted to recover the total interaction energy between them\cite{paper1}. The results of this method of determining interaction energies is shown in Fig. \ref{fig5}, together with the corresponding results derived in the previous section. It can be seen that as the number of masses increases, the total interaction energy increases to the same magnitude as the total proper mass in the universe (but with the opposite sign).
\vspace{-5pt}

\vspace{-25pt}
\section{Cosmological Consequences of Interaction Energies}

\begin{figure}[b!]
\begin{center}
  \parbox{2.5in}{\includegraphics[width=2.5in]{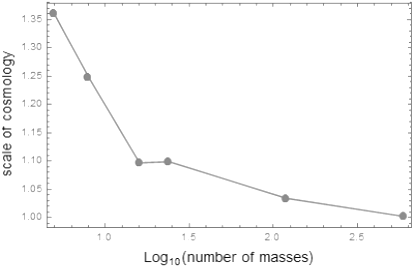}\figsubcap{a}}
  \hspace{-7pt}
  \parbox{2.5in}{\includegraphics[width=2.5in]{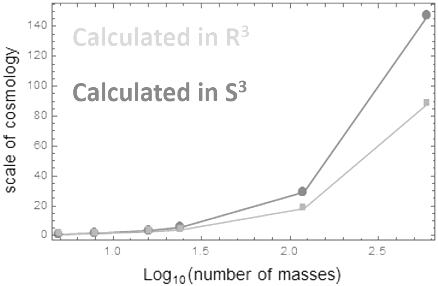}\figsubcap{b}}
  \caption{Plots showing (a) the scale of the cosmological models when the same total mass is divided between different numbers of black hole, and (b) the scale of cosmology when this is done with the same total energy (mass plus interactions).}
  \label{fig6}
\end{center}
\end{figure}

When the masses in these models are identical and regularly distributed, then the difference in scale of between models with the same total proper mass can be calculated\cite{paper1}. This is shown graphically in Fig. \ref{fig6}(a), where the scale of the cosmology has been normalised here by the scale of a dust-filled Friedmann solution (with the same total mass). It can be seen that as the mass is divided between larger numbers of black holes, the scale of the cosmology decreases. It can also be shown that as the number of masses diverges, the scale approaches the Friedmann value\cite{kor}. Unlike in the case of the cluster, interactions over large scales do not seem to contribute anything to the scale of the cosmology.

However, if we distribute the same amount of energy between different numbers of black holes, but this time keep the same total energy (mass plus interaction), then the results are quite different\cite{paper1}. This is shown graphically in Fig. \ref{fig6}(b), for the two methods of calculating interaction energies discussed above. Now, unlike the case where the interaction energies are ignored (and also unlike the asymptotically flat case), redistributing masses seems to result in a considerable change in the large-scale properties of the space: Dividing the same total energy between more black holes dramatically increases the scale of the cosmology.

\section{Conclusions}

We have investigated the problem of how interaction energies gravitate in cosmology, and whether or not they can contribute to the large-scale properties of a cosmological model. We have found that the interactions between nearby masses in a cluster do gravitate, and that they contribute to the large-scale cosmology in almost exactly the same way as the energy in mass. We have also shown that there is good reason to think that the interaction energies between masses separated by cosmological distances is large (although there are difficulties in defining them in this case). While large, however, it seems that interactions between bodies separated by cosmological scales do not contribute substantially to the scale of the universe.

These results beg the following question: If interactions on cosmological scales are large, and they do not contribute to the scale of the cosmology, then what happens when structure forms, and the magnitude of the interactions changes? This answer to this question is not yet known.

\vspace{10pt}
\flushleft
{\bf Acknowledgments.} I am grateful to Daniele Gregoris, Kjell Rosquist and Reza Tavakol, who were my collaborators in the scientific work on which this article is based.

\end{document}